\documentclass[12pt,dvips]{article}
\usepackage[dvips]{epsfig}

\usepackage{theorem}
\usepackage{defines}
\usepackage{maps}

\usepackage{amsfonts}

 1

\begin{document}

\date{September 1996}
\title{
\vspace{-5.0cm}
\begin{flushright}
{\normalsize UNIGRAZ-}\\
\vspace{-0.3cm}
{\normalsize UTP-}\\
\vspace{-0.3cm}
{\normalsize 17-09-96}\\
\end{flushright}
\vspace*{2.5cm}
Classical $\U1$ Lattice Gauge Theory in $D=2$}

\author
{\bf H. Gausterer and M. Sammer
 \\  \\
{\normalsize Institut f\"ur Theoretische Physik}\\
{\normalsize Universit\"at Graz}\\
{\normalsize A-8010 Graz, AUSTRIA} }

\maketitle

\begin{abstract}

Under the hypothesis of no topological structure below a certain scale,
we prove that any $\U1$ lattice configuration corresponds to a
classical $\U1$ gauge field with zero local field strength; i.e. any
local representative of the pullback connection one-form is a pure
gauge and the local curvature two-form is thus identical zero.  The topological
information is completely carried by the chart transitions. To each such
$\U1$ lattice configuration we assign a Chern number, which generally
depends on the reconstruction of the bundle and is only unique under
certain restrictions.

\end{abstract}

\thispagestyle{empty}

\newpage

\section{Motivation}

	There is a recent increased interest in $QED_2$. This concerns
	the continuum as well as the lattice version of the model
	(c.f. \cite{Fr92}
	\nocite{Fr93,GaSe94,GaLa94b,AzDiGa94,Fr95,GaLa95,Ga95,Ga95a,AzDiGa96b,IrSe96}-
	\cite{IrSe96}).
	The one flavor massless continuum model \cite{Sc62a,Sc62b} is
	analytically solvable and has been studied extensively. The
	reason for the increased interest is that $QED_2$ shows $QCD_4$
	like behavior.  This applies especially to the multi flavor
	situation \cite{GaSe94}.  The Maxwell equations for two
	dimensional electrodynamics also have topologically non trivial
	$C^\infty$ solutions with finite action which can be classified
	by their Chern number. These topological objects called
	instantons are considerably simpler to imagine for $\U1$ in
	$D=2$ than for $SU(2)$ in $D=4$ which is an additional appeal
	to study $QED_2$.  Therefore one finds in $QED_2$ three
	closely related problems.  There is the problem of the
	$\theta$-vacua, which naively speaking are superpositions of
	all topological sectors corresponding to different Chern
	numbers.  Also observed in both models is the occurrence of the
	${\rm U}_A(1)$ problem \cite{Ad69,BeJa69}. $QED_2$ further
	allows for a Witten-Veneziano type formula
	\cite{Wi79,Ve79,GaSe94}.

	It is not clear how important these topological nontrivial
	configurations are indeed for quantum physics. Naively such
	$C^\infty \in \cal{S}^\ast$ solutions should not contribute in
	the functional integration since the subset of such smooth
	solutions is of measure zero for the measure over
	$\cal{S}^\ast$.  Nevertheless the topological susceptibility,
	which is the first Chern character for $QED_2$ vice versa the
	second Chern character for $QCD_4$, appears in the ${\rm
	U}_A(1)$ anomaly.

	The lattice situation is quite different. First of all the
	lattice regularized version is analytically not solvable.
	Further assuming that the lattice model approximates in a
	certain limit the continuum model and thus also contributions
	from topology it is a priori not clear what differential
	geometry means for a set of points. Any straightforward bundle
	reconstruction will only lead to trivial bundles with Chern
	number zero. One way out is to provide the lattice with a very
	special topology and construct partially ordered sets which
	allow for non trivial bundles \cite{BaAl96}. $QED_2$ can be
	also defined on a fuzzy sphere which allows a topological
	classification in a surprisingly intuitive way via the Hopf
	fibration \cite{GrMa92,GrKl96}. A third possibility is to
	regard the lattice as a directed complex with a certain
	realization like $\torus^2$. This idea was pioneered by
	L\"uscher \cite{Lu82a} for $SU(2)$ in $D=4$ and put on a more
	axiomatic approach in \cite{Ph84} for $\U1$ in $D=2$.

	Without the explicit construction of bundles the
	$\theta$-vacuum problem and the topological charge problem on
	the lattice could also be addressed by possible remnants of the
	Atiyah Singer index theorem \cite{AtSi68}. For the numerical
	simulation of these models it turns out that {\it lattice
	topological charge} \cite{SeSt82} leads to an unpleasant
	problem.  As observed by \cite{SmVi87, SmVi87a, Vi88} the
	lattice Dirac operator indeed shows (approximate) zero modes
	depending on the {\it lattice topological charge} of the
	configuration.  The lattice Dirac operator thus cannot be
	inverted and the numerical procedure breaks down for such
	configurations, although the measure of the configuration is
	almost zero.

	In this paper we follow the strategy pioneered by L\"uscher
	\cite{Lu82a} and assume that the lattice is a directed 
	two-complex with $\torus^2$ as realization.  We further assume that
	the topological structure is trivial below a certain scale
	(i.e. within a region which is about of the size of a
	plaquette). This means, that any local pullback connection
	one-form is a pure gauge. This assumption is physically
	justified, since in the continuum limit it is assumed  that any
	local lattice structure does not contribute. Formally it
	shrinks to a point and thus has no structure.

\section{Classical Lattice Gauge Theory}

	Let us introduce the concept of a classical lattice model which
	is used to approximate classical gauge theory.

\begin{Definition}
	Let $\Lambda$ be a $2$-d complex and $\BB$ be a realization of 
	$\Lambda$, i.e. the space underlying the complex $\Lambda$.
	The complex $\Lambda$ 
	is called \EM{lattice on $\BB$}.
	A $0$-cell $x$ of $\Lambda$ is called \EM{site} and a 
	directed $1$-cell $\link{xy}$ of $\Lambda$ is called \EM{link} or
	\EM{bond}.
\end{Definition}

\begin{Definition}
	Let $\xi=(\EE,\pi,\BB)$ be a principal $G$ bundle, 
	$\omega:\TT\EE\to \GG$ be a connection one-form and
	$\Lambda$ be a lattice on $\BB$.
	The bundle $\xi_\Lambda=(\EE_\Lambda,\pi_\Lambda,\Lambda)$ 
	is called
	\EM{lattice-bundle} and
	the tuple $(\xi_\Lambda,\omega)$ is called
	\EM{classical lattice model}.
\end{Definition}

	Let $j:\Lambda\to \BB$ be the inclusion map.
	Then the lattice bundle $\xi_\Lambda$ could be identified with
	the restriction $\xi\restrict{\Lambda}$.
	The induced bundle $j^\ast(\xi)$ of $j$ is the bundle
	$(\EE',\pi',\Lambda)$ with the total-space
	$$\EE'=\{(x,e)\in\Lambda\times\EE \restr j(x)=\pi(e)\}$$
	 and the projection
	$\pi'=\pr_1$.
	On the other hand we have
	an isomorphism $(u, \id_\Lambda)$ to the induced bundle
	$j^\ast(\xi)$, i.e. the following diagram commutes
	\diagramquad[\EE_\Lambda-\pi_\Lambda->\Lambda-\id_\Lambda->\BB'<-\pi'-\EE'<-u-]
	with $u:\EE_\Lambda\to \EE':x\mapsto (\pi_\Lambda(e),e)$ and
	$e\in\EE$.
	Finally one obtains the following commutative diagram:
	$$ \begin{array}{rcccl}
	\EE_\Lambda & \maprightt{u} & \EE' & \maprightt{\hat j} & \EE\\
	\mapdownr{\pi_\Lambda} & & \mapdownr{\pi'} & & \mapdownr{\pi}\\
	\Lambda & \maprightt{\id_\Lambda} & \BB' & \maprightt{j} & \BB
	\end{array} $$ 
	where $\hat j$ is defined as usual by  $\hat j:\EE'\to \EE:(x,e)\mapsto e$. 
	We also know that each fiber of the pullback $j^\ast(\xi)$ is
	homeomorphic to the fiber $G$ of $\xi$.
	Therefore our lattice bundle $\xi_\Lambda$ has typical fiber $G$
	and is also a principal $G$-bundle.

\begin{Definition}
	Let  
	$\Lambda$ be a lattice on $\BB$ and $x_0, x_1$ two neighboring 
        $0$-cells.
	$\gamma:[0,1] \to\BB:0\mapsto x_0:1\mapsto x_1$ be a path in $\BB$.
	The corresponding image in $\Lambda$ is the
	directed $1$-cell $\link{x_0 x_1}$, 
	and called \EM{path in $\Lambda$}.
\end{Definition}	

	If the path $\gamma$ is a \EM{loop} then the corresponding path in
	$\Lambda$ is a $1$-cycle.

\begin{figure}
	\begin{center}	
		\psfig{figure=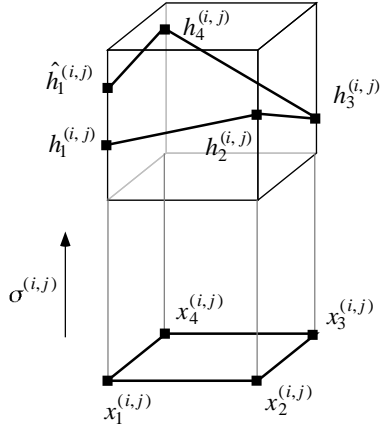}
	\end{center}
		\caption{Path of the horizontal lift $\tilde\gamma$ 
		of $\gamma$ in $\EE_\Lambda$}
		\protect\label{f-lat-trans}
\end{figure}

\begin{Definition}\label{d-lat-parallel}	
	Let $(\xi_\Lambda,\omega)$ be a 	
	classical lattice model $\gamma:[0,1] \to\BB$ be a path and $\link{x_0 x_1}$ be 
	the corresponding path in $\Lambda$.
	The \EM{lattice parallel translation} along the path $\gamma$ is
	a map
	$$\transport_{\link{x_0 x_1}}:\pi_\Lambda\inv(x_0)\to
	\pi_\Lambda\inv(x_1):h_0\mapsto h_1$$
	where $h_1$ denotes the parallel transport of $h_0$ along
	the horizontal lift $\tilde\gamma$ of $\gamma$, i.e.
	$\tilde\gamma(0)=h_0$ and $h_1:= \tilde\gamma(1)$.
\end{Definition}

	Let $(\xi_\Lambda,\omega)$ be a classical lattice model,
	$\sigma:U\to\EE_\Lambda$ be a local section.  One obtains the
	lattice parallel translation $\transport_{\link{x_0 x_1}}$ in
	terms of the local connection one-form $\ga = \sigma^\ast
	\omega$ \begin{equation}\label{e-local-lat-trans}
	\transport_{\link{x_0 x_1}}:h_0\mapsto h_1=h_0 \circ {\bf P}
	\exp\left( -\int_{x_0}^{x_1} j^\ast\ga \right), \end{equation}
	where the boundary condition of the horizontal lift function

	$$g(t)={\bf P} \exp\left(
	-\int_{x(0)}^{x(t)} j^\ast\ga \right), $$ has been set to
	$g(0):=e$.

\begin{Definition}
	Let $(\xi_\Lambda,\omega)$ be a 	
	classical lattice model.
	To each $1$-cell one can assign a lattice parallel translation
	which leads to a map
	$$\link{xy}\mapsto \transport_{\link{xy}}$$
	which is called a \EM{gauge field on $\Lambda$}. 
	The collection $\{\transport_{\link{xy}}\}$ of all this lattice parallel
	translations
	is called \EM{configuration on $\Lambda$}.	
\end{Definition}

	In general one cannot define a global gauge field
	on $\Lambda$ except the bundle $\xi_\Lambda$ is a trivial bundle.
	Therefore a configuration contains elements which belong to
	different local trivialisations.

\begin{Definition}
	Let 
	$\Lambda$ be a complex such that the realization of $\Lambda$ is the
	2-Torus $\torus^2$.
	A directed complex $\Lambda$ with
	\begin{enumerate}
		\item $0$-cells $\lpoint{i}{j}$,
		\item $1$-cells $(i,i+1)\times \{j\}$ and
		 $\{i\}\times (j,j+1)$
		\item $2$-cells $(i,i+1)\times (j,j+1)$
	\end{enumerate}
	for all $i\in\ZZ_M$ and $j\in\ZZ_N$
	is called a \EM{cubic lattice on $\torus^2$} and
	is denoted by $\Lambda(N,M)$.	
	The closure of a $2$-cell $(i,i+1)\times (j,j+1)$ is called \EM{plaquette}
	and is denoted $\Lambda_{(i,j)}$. 
\end{Definition}

	Since the 2-Torus $\torus^2$ cannot be covered
	by a single chart
	we choose an atlas 
	$${\cal A}(\torus^2):=\{
	(U_{(i,j)},\phi_{(i,j)})\restr 0\leq i\leq M-1, 0\leq j\leq
	N-1\}$$
	where the charts be
	all the open subsets
	$U_{(i,j)}\subset \torus^2$ which
	cover the corresponding $2$-cells $(i,i+1)\times (j,j+1)$.
	\begin{figure}
	\begin{center}	
		\psfig{figure=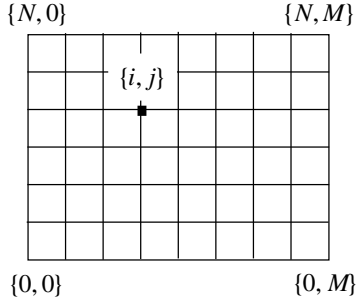}
	\end{center}
		\caption{Cubic lattice on $\torus^2$}
		\protect\label{f-lattice}
	\end{figure}

	Let $U_{(i,j)}$ be a chart on $\torus^2$.
	We denote the corresponding local section/trivialisation
	by $\sigma^{(i,j)}(x)=\vphi^{(i,j)}(x,g^{(i,j)})$
	and $\vphi^{(i,j)}$, respectively.
	The local connection one-form is denoted by $\ga^{(i,j)}$.
	Since we denote an open interval by $(i,j)$ 
	a site is denoted by $\lpoint{i}{j}\in\torus^2$.

	To make the lattice bundle $\xi_\Lambda$ unique
	one has to fix the collection
	of all transition functions
	$\{ t_{(i,j)(k,l)}(x) \}$.
	Our goal is to reconstruct the transition functions, i.e. lattice bundle, from
	a given configuration of the lattice model.

	In order to define our $\U1$ gauge theory over $\torus^2$ one
	needs to specify a {\em global} connection one-form
	$$\omega:\TT\EE\to\I\RR.$$
	Since we are interested in a connection form which has a
	trivial topological structure in a local trivialisation
	$\vphi^{(i,j)}$ (no topological structure below a certain scale)
	we define the {\em local} connection one-forms
	to be
	\begin{equation}\label{e-localconnection}
	\ga^{(i,j)}\restrict{U_{(i,j)}}=\sigma^{(i,j)\ast} 
	\omega:={t^{(i,j)}}\inv (p) \circ\baseCTB{t^{(i,j)}}(p),
	\end{equation}	
	for all $p\in U_{(i,j)}$,
	i.e. the local connection one-form restricted to the chart $U_{(i,j)}$
	has to be a pure gauge in the local trivialisation
	$\vphi^{(i,j)}$.
	This connection together with the lattice bundle
	$\xi_{\Lambda(M,N)}=(\EE_\Lambda,\pi_\Lambda,\Lambda(M,N))$
	defines our model $(\xi_{\Lambda(M,N)}, \omega)$.

	Since the choice of all the $g^{(i,j)}$ is arbitrary this leads
	to $N\cdot M$ degrees of freedom.
	The choice of the $g^{(i,j)}$ is equivalent to the
	choice of the local trivialisations $\vphi^{(i,j)}$, 
	but due to left invariance of our connection one-form (Cartan Maurer form)
	the final result does not depend on these degrees.

\section{Reconstruction of the Bundle}

	This property of the connection one-form $\omega$ leads to some
	restrictions in the choice of local trivialisations.  In
	general, the only information one has are the 'transporters'
	which are assigned to each link of the lattice, i.e. the
	configuration of the lattice model.  Since we have an atlas
	${\cal A}(\torus^2)$ of the torus one has to be careful how to
	assign the 'transporters' to the given charts.

\begin{Lemma}\label{l-charts-choice}
	Let $(\xi_{\Lambda(M,N)}, \omega)$ be our lattice model.
	Let $U_{(i,j)}$ be a chart on $\torus^2$ and $\Lambda_{(i,j)}$
	the corresponding plaquette. Let 
	${\cal A}(\torus^2)$
	be our atlas of $\torus^2$ and 
	$$\omega^{(i,j)}\restrict{U_{(i,j)}}=\sigma^{(i,j)\ast} 
	\omega:={t^{(i,j)}}\inv(p)
	\circ\baseCTB{t^{(i,j)}}(p)$$
	our local connection
	one-form.
	Let $\{\transport\}$ be a configuration.
	Only three of the four lattice parallel translations
	$$\transport^{(i,j)}_{\link{x_1 x_2}},\, \transport^{(i,j)}_{\link{x_2
	x_3}},\,\transport^{(i,j)}_{\link{x_3 x_4}}\,\mbox{and}\,\,\,
	\transport^{(i,j)}_{\link{x_4 x_1}}$$
	which belong to the plaquette $\Lambda_{(i,j)}$ can be assigned to
	the corresponding local trivialisation 
	$\vphi^{(i,j)}$, i.e. belong to the same local representation.
\end{Lemma}
\begin{proof}
	Since the local connection one-form $\omega^{(i,j)}\restrict{U_{(i,j)}}$
	is a pure gauge the lattice parallel translations around the plaquette
	must be closed. 
	Therefore the lattice parallel
	translation $\transport^{(i,j)}_{\gamma}$ has to be the group
	identity $e$, thus
	three of the four lattice parallel translations
	have to be given in the local trivialisation and the fourth
	has to be the inverse of the composition of the given three.	
\end{proof}

	The next step is to reconstruct the transition functions $\{
	t_{(i,j)(k,l)}(x) \}$ from a given configuration of the lattice
	model.

	\begin{figure}
	\begin{center}	
		\psfig{figure=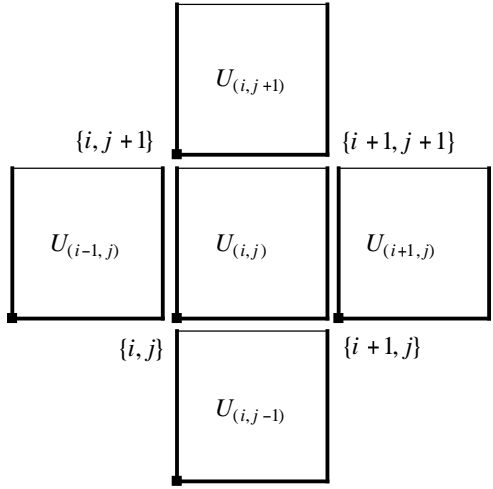}
	\end{center}
		\caption{Choice of the charts}
		\protect\label{f-t2charts}
	\end{figure}
	Take a local section $\sigma^{(i,j)}$ together with the four
	neighboring local sections $\sigma^{(i-1,j)}$,
	$\sigma^{(i+1,j)}$, $\sigma^{(i,j-1)}$ and $\sigma^{(i,j+1)}$.

	Denote the transition function which maps from the fiber $\U1$
	in the local trivialisation $\vphi^{(i-1,j)}$ to the same fiber
	in the local trivialisation $\vphi^{(i,j)}$ at $\lpoint{k}{l}$
	by 
	$$t_{(i,j)(i-1,j)}(\lpoint{k}{l}),$$
	we obtain the following relation for the elements
	$h^{(i-1,j)}(\lpoint{k}{l})$ and $h^{(i,j)}(\lpoint{k}{l})$ of
	$\U1$:
	\begin{equation}\label{e-sect-trans}
		h^{(i-1,j)}(\lpoint{k}{l})=h^{(i,j)}(\lpoint{k}{l})
		\circ t_{(i,j)(i-1,j)}(\lpoint{k}{l}). 
	\end{equation}
	Since we want to calculate the transition function from the
	local sections we rewrite (\ref{e-sect-trans}) to obtain
	\begin{equation}\label{e-trans-sect}
		t_{(i,j)(i-1,j)}(\lpoint{k}{l})=
		{h^{(i,j)}}\inv(\lpoint{k}{l}) \circ
		h^{(i-1,j)}(\lpoint{k}{l}).  
	\end{equation}

	In each local
	trivialisation $\vphi^{(i,j)}$ the local connection one-form
	$\omega^{(i,j)}\restrict{U_{(i,j)}}$ has to be a pure gauge.

	We choose our charts according to Fig.~\ref{f-t2charts} where
	the three links which correspond to the three
	lattice parallel translation which are assigned to
	the corresponding local trivialisation 
	$\vphi^{(k,l)}$ are marked as bold lines. 

	In a trivialisation $\vphi^{(i,j)}$ we can
	express the lattice parallel translation
	in terms of the local connection one-form $\ga^{(i,j)}$ by
	\begin{equation}\label{e-t2local-lat-trans}
	\transport^{(i,j)}_{\link{x_1 x_2}}:h_1^{(i,j)}\mapsto
	 h_2^{(i,j)} = h_1^{(i,j)}\circ {\bf P}\exp\left(
	-\int_{x_0}^{x_1} j^\ast \ga^{(i,j)} \right).
	\end{equation}

	Since we have one degree of freedom per local
	trivialisation we choose 
	$$h_1^{(i,j)}:=g^{(i,j)}$$
	where $g^{(i,j)}$ is an arbitrary $\U1$-element.
	
	Denote the three lattice parallel translations along the links 
	$\link{x_1 x_2}$, $\link{x_2 x_3}$ and $\link{x_1 x_4}$ by
	$\transport^{(i,j)}_{\link{x_1 x_2}}, \transport^{(i,j)}_{\link{x_2
	x_3}}$ and
	$\transport^{(i,j)}_{\link{x_1 x_4}}$, respectively.	
	The fourth
	lattice parallel translation is nothing but
	$$\transport^{(i,j)}_{\link{x_3 x_4}}:=\transport^{(i,j)}_{\link{x_3
	x_2}}\circ
	\transport^{(i,j)}_{\link{x_2 x_1}}  
	\circ\transport^{(i,j)}_{\link{x_1 x_4}},$$
	since our local connection one-form has to be a pure gauge.
	\begin{table}
		\caption{Notation of local coordinates $x$ and fiber elements $h$ in different charts}
		\begin{tabular}{llllll}
		\noalign{\smallskip\hrule\smallskip}
		point of $\torus^2$ & $U_{(i,j)}$ & $U_{(i-1,j)}$ & $U_{(i+1,j)}$
		& $U_{(i,j-1)}$ & $U_{(i,j+1)}$\\
		\noalign{\smallskip\hrule\smallskip}		
		$\lpoint{i}{j}$ & $x_1^{(i,j)}$ & $x_2^{(i-1,j)}$ & - & $x_4^{(i,j-1)}$ & - \\
		$\lpoint{i+1}{j}$ & $x_2^{(i,j)}$& - & $x_1^{(i+1,j)}$ & $x_3^{(i,j-1)}$ & - \\
		$\lpoint{i+1}{j+1}$ & $x_3^{(i,j)}$& - & $x_4^{(i+1,j)}$ & - & $x_2^{(i,j+1)}$ \\	
		$\lpoint{i}{j+1}$ & $x_4^{(i,j)}$& $x_3^{(i-1,j)}$ & - & - & $x_1^{(i,j+1)}$ \\[3ex]
		$\lpoint{i}{j}$ & $h_1^{(i,j)}$ & $h_2^{(i-1,j)}$ & - & $h_4^{(i,j-1)}$ & - \\
		$\lpoint{i+1}{j}$ & $h_2^{(i,j)}$& - & $h_1^{(i+1,j)}$ & $h_3^{(i,j-1)}$ & - \\
		$\lpoint{i+1}{j+1}$ & $h_3^{(i,j)}$& - & $h_4^{(i+1,j)}$ & - & $h_2^{(i,j+1)}$ \\	
		$\lpoint{i}{j+1}$ & $h_4^{(i,j)}$& $h_3^{(i-1,j)}$ & - & - & $h_1^{(i,j+1)}$ \\				
		\noalign{\smallskip\hrule\smallskip}
		\end{tabular}
		\protect\label{t-t2notat}
	\end{table}		
	We 'transport' the element $g^{(i,j)}$ at $x_1^{(i,j)}$ via
	these lattice parallel translations to obtain the fiber elements
	at all sites (c.f. Fig.~\ref{f-lat-trans}) of this plaquette:
	$$
	\begin{array}{l} 
	h_1^{(i,j)}:=g^{(i,j)},\\
	h_2^{(i,j)}=g^{(i,j)} \circ\transport^{(i,j)}_{\link{x_1 x_2}},\\
	h_3^{(i,j)}=g^{(i,j)} \circ\transport^{(i,j)}_{\link{x_1 x_2}}  
		\circ\transport^{(i,j)}_{\link{x_2 x_3}},\\
	h_4^{(i,j)}=g^{(i,j)} \circ\transport^{(i,j)}_{\link{x_1 x_4}}.\\
	\end{array}
	$$

	Now we calculate the transition functions from the 
	local trivialisations $\vphi^{(i,j)}$.
	\begin{figure}
	\begin{center}	
		\psfig{figure=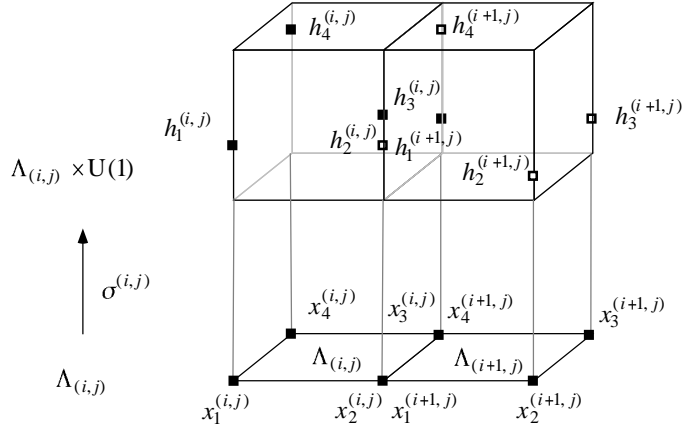}
	\end{center}
		\caption{Notation of the local sections}
		\protect\label{f-t2notation}
	\end{figure}
	Each site is covered by four charts.
	The first step is to recognize that only three
	of the four transition functions have to be calculated since
	the cocycle conditions give
	some additional relations.
	
	We use the charts according to Fig.~\ref{f-t2charts} and
	summarize the notation of the local coordinates in Table~\ref{t-t2notat}.

	Our choice of charts gives the two relations
	\begin{equation}\label{e-simpl}
	\transport^{(i,j)}_{\link{x_3 x_2}}=\transport^{(i+1,j)}_{\link{x_4 x_1}}
	\quad\mbox{and}\quad \transport^{(i,j)}_{\link{x_4 x_1}}=
	\transport^{(i-1,j)}_{\link{x_3 x_2}},
	\end{equation}
	which can be used to simplify the results.
	Also in the non-Abelian case they are useful because if one
	calculates Chern classes one takes the trace over the
	transition functions.

	For the Abelian case together with 
	the two relations of (\ref{e-simpl}) and
	with the use of
	(\ref{e-trans-sect})
	we obtain:
	\begin{itemize}
	\item Site $\lpoint{i}{j}$
		\begin{eqnarray}\label{trans1} 
		t_{(i,j)(i-1,j)}(\lpoint{i}{j}) & = & {g^{(i,j)}}\inv \circ 
			g^{(i-1,j)} \circ\transport^{(i-1,j)}_{\link{x_1x_2}}	\nonumber \\ 
		t_{(i,j)(i,j-1)}(\lpoint{i}{j}) & = & {g^{(i,j)}}\inv \circ 
			g^{(i,j-1)} \circ\transport^{(i,j-1)}_{\link{x_1 x_4}}
		\end{eqnarray}
	\item Site $\lpoint{i+1}{j}$
		\begin{eqnarray}\label{trans2} 	
		t_{(i,j)(i,j-1)}(\lpoint{i+1}{j}) & = & \transport^{(i,j)}_{\link{x_2 x_1}}\circ
			{g^{(i,j)}}\inv \circ 
			g^{(i,j-1)} \circ\transport^{(i,j-1)}_{\link{x_1 x_2}}  
			\circ\transport^{(i,j-1)}_{\link{x_2 x_3}} \nonumber \\
		t_{(i,j)(i+1,j)}(\lpoint{i+1}{j}) & = &  \transport^{(i,j)}_{\link{x_2 x_1}}\circ
			{g^{(i,j)}}\inv \circ g^{(i+1,j)}
		\end{eqnarray}
	\item Site $\lpoint{i+1}{j+1}$
		\begin{eqnarray}\label{trans3} 
		t_{(i,j)(i+1,j)}(\lpoint{i+1}{j+1}) & = & 
			\transport^{(i,j)}_{\link{x_2 x_1}} \circ
			{g^{(i,j)}}\inv \circ 
			g^{(i+1,j)} \\	\nonumber		
		t_{(i,j)(i,j+1)}(\lpoint{i+1}{j+1}) & = &\transport^{(i,j)}_{\link{x_3 x_2}} \circ
			\transport^{(i,j)}_{\link{x_2 x_1}} \circ
			{g^{(i,j)}}\inv \circ 
			g^{(i,j+1)} \circ\transport^{(i,j+1)}_{\link{x_1 x_2}}
		\end{eqnarray}
	\item Site $\lpoint{i}{j+1}$ 	
		\begin{eqnarray}\label{trans4} 
		t_{(i,j)(i,j+1)}(\lpoint{i}{j+1}) & = & \transport^{(i,j)}_{\link{x_4 x_1}}\circ
			{g^{(i,j)}}\inv \circ g^{(i,j+1)} \nonumber \\
		t_{(i,j)(i-1,j)}(\lpoint{i}{j+1}) & = &
			{g^{(i,j)}}\inv \circ 
			g^{(i-1,j)} \circ\transport^{(i-1,j)}_{\link{x_1 x_2}}  
		\end{eqnarray}
	\end{itemize}

\section{Topological Invariants}
	
The Chern character is used to measure the twist of
a bundle.
Integrating the first Chern character $\chch_1(\gf)$
over the whole lattice gives an integer called
\EM{Chern number}
$$\Ch(\xi_{\Lambda(M,N)}):=\int_{\torus^2}\chch_1(\gf)=
	\frac{\I}{2\pi}\int_{\torus^2}\gf,$$
which is a topological invariant and which can be used
to classify the $\U1$-bundles over $\Lambda(M,N)$.

	One has to be careful if integrating over $\Lambda(M,N)$ since
	our bundle is constructed by patching together local pieces
	via the transition functions.
	One also should remember that integration of a $n$-form over
	a manifold is done via integration over
	$n$-cells in the  corresponding complex. 
	Let $\ga$ be a $2$-form and $j:\Lambda(M,N) \to \torus^2$.
	Then one writes simply
	$$\int_{\torus^2}\ga$$
	for
	$$\int_{\torus^2}\ga:=\int_{\Lambda(M,N)}j^\ast\ga,$$
	because the integral is independent of the cellular subdivision.

	Let $\{\lambda_{(i,j)}\}$ be a partition of unity subordinate to
	the covering $\{U_{(i,j)}\}$.
	Then our pullback {\em global} connection one-form
	can be written as 
	\begin{equation}\label{e-pb-conn}
		\ga:=\sum_{(i,j)\in\ZZ_M\times\ZZ_N} 
		\lambda_{(i,j)}\,\ga=\sum_{(i,j)\in\ZZ_M\times\ZZ_N} \ga_{(i,j)}.
	\end{equation}	
	Therefore we get 
	$$\gf=\baseCTB{}\,\ga=\sum_{(i,j)\in\ZZ_M\times\ZZ_N}
		\baseCTB{}\,\ga_{(i,j)}.$$
	Integration is now be done via partition of unity by
	$$\int_{\torus^2}\gf:=\sum_{(i,j)\in\ZZ_M\times\ZZ_N} 
		\int_{U_{(i,j)}}\baseCTB{}\,\ga_{(i,j)}.$$
	Since our lattice model $(\xi_{\Lambda(M,N)},\omega)$
	is designed in such a way that
	there is 
	no topological structure below a certain scale
	we have
	$$
	\ga^{(i,j)}:=\ga^{(i,j)}\restrict{U_{(i,j)}}=\sigma^{(i,j)\ast} 
	\omega:={g^{(i,j)}}\inv (x) \circ\baseCTB{g^{(i,j)}}(x),
	$$	
	for all $x\in U_{(i,j)}$.	
	We notice that the part of our pullback {\em global} connection one-form
	with compact support on $U_{(i,j)}$ denoted by $\ga_{(i,j)}$
	is obtained by
	rewriting
	$$
	\ga\restrict{U_{(i,j)}}=\ga_{(i,j)}+
		\sum_{\mbox{\scriptsize neighbors}} \ga_{(k,l)}\restrict{U_{(i,j)}
		\cap U_{(k,l)}}.
	$$
	to get
	$$
	\ga_{(i,j)}=\ga\restrict{U_{(i,j)}}-
	\sum_{\mbox{\scriptsize neighbors}} \ga_{(k,l)}\restrict{U_{(i,j)}
	\cap U_{(k,l)}}	
	.$$		

	\begin{figure}
	\begin{center}	
		\psfig{figure=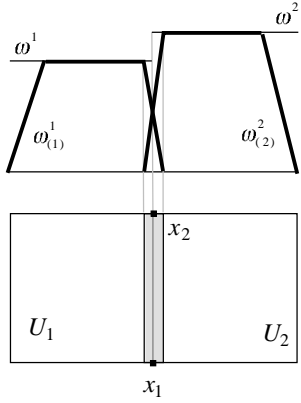}
	\end{center}
		\caption{Partition of the connection one-form}
		\protect\label{f-t2ovl}
	\end{figure}
	Let $(\xi_{\Lambda(M,N)},\omega)$ be 
	our lattice model. Take overlapping charts
	$U_1$ and $U_2$ on $\torus^2$ and let $\ga^1$ and $\ga^2$ be
	the local connection one-form on $U_1$ and $U_2$, respectively.
	Let $\{\lambda_{(i)}\}$ be a partition of unity subordinate to
	the covering $\{U_i\}$.
	The corresponding pullback connection one-form is
	$\ga\restrict{U_1\cup U_2}=\ga_{(1)}+\ga_{(2)}$. With
	the two relations
	$$\ga_{(1)}=\ga\restrict{U_1}-\ga_{(2)}\restrict{U_1\cap U_2}$$
	and
	$$\ga_{(2)}=\ga\restrict{U_2}-\ga_{(1)}\restrict{U_1\cap U_2}$$
	the integral
	\begin{eqnarray*}
		\int_{U_1\cup U_2}\baseCTB{}\,\ga & = &
		\int_{U_1}\baseCTB{}\,\ga_{(1)}+\int_{U_2}\baseCTB{}\,\ga_{(2)}\\
		& = & \int_{U_1}\baseCTB{}\,\ga\restrict{U_1} -
		\int_{U_1}\baseCTB{}\,\ga_{(2)}\restrict{U_1\cap U_2}\\
		& & + \int_{U_2}\baseCTB{}\,\ga\restrict{U_2} -
		\int_{U_2}\baseCTB{}\,\ga_{(1)}\restrict{U_1\cap U_2}
	\end{eqnarray*}
	expands to
	$$\int_{U_1\cup U_2}\baseCTB{}\,\ga=-\int_{U_1}\baseCTB{}\,\ga_{(2)}\restrict{U_1\cap U_2}
		-\int_{U_2}\baseCTB{}\,\ga_{(1)}\restrict{U_1\cap U_2},$$
	where we had assumed that the local connection forms
	have to be pure gauges, i.e.
	$\baseCTB{}\,\ga^1\restrict{U_1}\equiv 0$ and
	$\baseCTB{}\,\ga^2\restrict{U_2}\equiv 0$.
	Applying Stokes' theorem gives
	$$\int_{U_1\cup U_2}\baseCTB{}\,\ga=-\int_{\partial U_1}
		\ga_{(2)}\restrict{U_1\cap U_2}	
		-\int_{\partial U_2}\ga_{(1)}\restrict{U_1\cap U_2}.$$
	Finally we realize (c.f. Fig~\ref{f-t2ovl}) that at the boundaries 
	of $U_1$ and $U_2$
	only the local connections $\ga^2$ and $\ga^1$, respectively, count.

	Note that due to the left invariance of our local connection one-form
	we have with $\tilde t(x) = g \circ t(x)$ and $g$ constant
	\begin{equation}\label{constg}
	t\inv(x)\circ\baseCTB{}\,t(x) = \tilde t\inv(x)\circ\baseCTB{}\,\tilde t(x).
	\end{equation}

	We further notice that due to the definition of the integral 
	over a cell-complex 
	 our map $j$ is an inclusion and can 
	be omitted. 
	Therefore we get
	$$\int_{U_1\cup U_2}\baseCTB{}\,\ga=-\int_{\link{x_1 x_2}}\ga^2
		-\int_{\link{x_2 x_1}}\ga^1,$$
	and together with
	$$\ga^1=\ga^2+
		t_{21}\inv \circ \,\baseCTB{}\,t_{21},$$
	the result
	\begin{eqnarray*}\label{e-int-trans}
		\int_{U_1\cup U_2}\baseCTB{}\,\ga &=& -\int_{\link{x_1 x_2}}\ga^2
		-\int_{\link{x_2 x_1}}\ga^2-\int_{\link{x_2 x_1}}
		t_{21}(x)\inv \circ \,\baseCTB{}\,t_{21}(x)\\
		& = & -\int_{\link{x_2 x_1}}
		t_{21}(x)\inv \circ \,\baseCTB{}\,t_{21}(x)\\
		& = & \log t_{21}(x_1)-\log  t_{21}(x_2).
	\end{eqnarray*}	
	
	If we further assume that
	\begin{equation}\label{lesspi}
        | \int_{U_1\cup U_2}\baseCTB{}\,\ga | < \pi
	\end{equation}
	then the above equation can be written as 
	\begin{equation}\label{e-chern2}
	\int_{U_1\cup U_2}\baseCTB{}\,\ga = \log \left( t_{21}(x_1) \circ
		t_{21}\inv(x_2) \right) ,
	\end{equation}
        where $\log ( t_{21}(x_1) \circ t_{21}\inv(x_2))$ is defined as
	the principal value with range $[-\pi,\pi)$.
	From (\ref{lesspi}) follows that $ t_{21}(x_1) \circ 
		t_{21}\inv(x_2)\not=-1$.   
	As we will see later there can be 
	\EM{configurations on $\Lambda$} which violate
	assumption (\ref{lesspi}). Since the values
	of each transition function $ t_{(i,j)(k,l)}(x) $ are only known
	on the two 
	end points of the region of integration, 
	a parameterization of $\U1$, such that at least 
	$$
	| \int_{U_1\cup U_2}\baseCTB{}\,\ga | \leq \pi
	$$
        holds, can always be assumed. Note that this assumption is an addition
	to (\ref{e-localconnection}). 

\begin{figure}
	\begin{center}	
		\psfig{figure=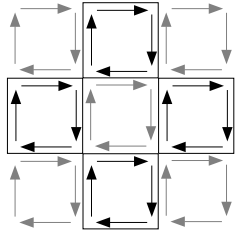}
	\end{center}
		\caption{Orientation of the plaquettes}
		\protect\label{f-t2orient}
	\end{figure}

	Due to the fact that
	on $U_{(i,j)}\cap U_{(k,l)}$ the local connection one-forms
	are related as
	$$\ga^{(k,l)}(x)=\ga^{(i,j)}(x) + 
		t_{(i,j)(k,l)}\inv(x)\circ\baseCTB{}\,t_{(i,j)(k,l)}(x) ,$$
	we obtain:
	\begin{eqnarray*} 
	\int_{\torus^2}\gf & = & - \sum_{\link{x_a x_b}}
		\int_{\link{x_a x_b}} t_{(i,j)(k,l)}(x)\inv \circ\baseCTB{}\,t_{(i,j)(k,l)}(x)\\
		& = & - \sum_{\link{x_a x_b}}\left[
		\log t_{(i,j)(k,l)}(x_b)- \log t_{(i,j)(k,l)}(x_a)\right],				
	\end{eqnarray*} 
	where the sum is over all directed links $\link{x_a x_b}$
	according to Fig.~\ref{f-t2orient}.

	Thus the Chern number is
	\begin{equation}\label{e-chernnumber}
		\Ch(\xi_{\Lambda(M,N)})=\frac{\I}{2\pi}
		\sum_{\link{x_a x_b}}\left[
		\log t_{(i,j)(k,l)}(x_a)- \log t_{(i,j)(k,l)}(x_b)\right].	
	\end{equation}

	When integrating over all links one should remember that our
	lattice is a directed complex, i.e. we have
	an orientation (c.f. Fig.~\ref{f-t2orient}).

	Let $M$ and $N$ be even
	integers,
	then the Chern number (c.f. (\ref{e-chernnumber})) gives
	\begin{eqnarray*}
		\Ch(\xi_{\Lambda}) & = &\frac{\I}{2\pi}
		\sum_{\{\bar\imath, \bar \jmath\}}\left[
		\log t_{(i,j)(i,j-1)}(\lpoint{i}{j})- \log
		t_{(i,j)(i,j-1)}(\lpoint{i+1}{j})\right]\\
		 & + & \frac{\I}{2\pi}
		\sum_{\{\bar\imath, \bar \jmath\}}\left[
		\log  t_{(i,j)(i,j+1)}(\lpoint{i+1}{j+1})- \log
		t_{(i,j)(i,j+1)}(\lpoint{i}{j+1})\right]\\
		 & + & \frac{\I}{2\pi}
		\sum_{\{\bar\imath, \bar \jmath\}}\left[
		\log t_{(i,j)(i-1,j)}(\lpoint{i}{j+1}) - \log
		t_{(i,j)(i-1,j)}(\lpoint{i}{j})\right]\\
		 & + & \frac{\I}{2\pi}
		\sum_{\{\bar\imath, \bar \jmath\}}\left[
		\log t_{(i,j)(i+1,j)}(\lpoint{i+1}{j+1}) - \log
		t_{(i,j)(i+1,j)}(\lpoint{i+1}{j})\right],
	\end{eqnarray*}
	where the sum is over all even or odd sites $\{\bar\imath, \bar \jmath\}$.
	The last two sums give zero because
	we have
	$$t_{(i,j)(i-1,j)}(\lpoint{i}{j})=t_{(i,j)(i-1,j)}(\lpoint{i}{j+1})$$
	and
	$$t_{(i,j)(i+1,j)}(\lpoint{i+1}{j})=t_{(i,j)(i+1,j)}(\lpoint{i+1}{j+1}).$$
	
	If we straightforwardly insert the transition functions then this 
	gives with the use of (\ref{constg})
	\begin{eqnarray}\label{e-ch-sum}
		\Ch(\xi_{\Lambda}) & = & \frac{\I}{2\pi}
		\sum_{\{\bar\imath, \bar \jmath\}}
			\log \transport^{(i,j-1)}_{\link{x_1 x_4}}
			-
			\log \transport^{(i,j)}_{\link{x_2 x_1}}\circ
			\transport^{(i,j-1)}_{\link{x_1 x_2}}  
			\circ\transport^{(i,j-1)}_{\link{x_2
			x_3}}\nonumber\\
		 & + & \frac{\I}{2\pi}
		\sum_{\{\bar\imath, \bar \jmath\}}
			 \log
			\transport^{(i,j)}_{\link{x_3 x_2}} \circ
			\transport^{(i,j)}_{\link{x_2 x_1}} \circ
			\transport^{(i,j+1)}_{\link{x_1 x_2}}
			-\log\transport^{(i,j)}_{\link{x_4 x_1}}
	\end{eqnarray}

	Note that this definition of the Chern number 
	is not lattice gauge invariant in the usual
	sense. This means that for a general 
	\EM{configuration on $\Lambda$}
	different lattice gauges might
	lead to different results for the Chern number.  We also note that
	reversing all transporters, which should lead to
	$-\Ch(\xi_{\Lambda})$, does in general not hold for the above result.
	To derive a unique result we must apply assumption (\ref{lesspi})
	and obtain
	\begin{eqnarray}\label{e-ch-sum1}
		\Ch(\xi_{\Lambda}) & = & \frac{\I}{2\pi}
		\sum_{\{\bar\imath, \bar \jmath\}}
		\log \Bigl
		( \transport^{(i,j-1)}_{\link{x_1 x_4}}
		\circ (\transport^{(i,j-1)}_{\link{x_1 x_2}}  
		\circ\transport^{(i,j-1)}_{\link{x_2 x_3}}
		\circ \transport^{(i,j)}_{\link{x_2 x_1}})\inv \Bigr) \nonumber\\
		 & + & \frac{\I}{2\pi}
		\sum_{\{\bar\imath, \bar \jmath\}}
		\log \Bigl( 
		\transport^{(i,j+1)}_{\link{x_1 x_2}} \circ
		\transport^{(i,j)}_{\link{x_3 x_2}} \circ
		\transport^{(i,j)}_{\link{x_2 x_1}} \circ
		(\transport^{(i,j)}_{\link{x_4 x_1}})\inv \Bigr)
	\end{eqnarray}
	In (\ref{e-ch-sum}) as well as in  (\ref{e-ch-sum1}) the sum 
	over all even sites can be replaced by the sum over 
	all odd sites replacing
	$(i,j)$ by $(i,j-1)$ and 
	$\log u$ by $-\log u\inv$. 
	Finally, we rewrite the second sum such that we can take the sum
	instead of all even sites over all sites $\lpoint{i}{j}$ and obtain
	the following theorem.

\begin{Theorem}\label{th-chern}

	Let $(\xi_{\Lambda(M,N)},\omega)$ be our lattice model and
	choose the charts according to Fig.~\ref{f-t2charts}.  The
	local connection one-form is a pure gauge and defined as in
	(\ref{e-localconnection}).  Let the transition functions be as
	in (\ref{trans1}) to (\ref{trans4}).  Assume that for each
	1-cell (link)
	$$|\int_{\link{x_a x_b}} t_{(i,j)(k,l)}(x)\inv
	\circ\baseCTB{}\,t_{(i,j)(k,l)}(x)| < \pi $$ 
	holds; i.e. for each 0-cell (site) $\{i,j\}$ we must have
	$$\transport^{(i,j)}_{\link{x_2 x_1}}
	\circ\transport^{(i,j-1)}_{\link{x_4 x_1}}
	\circ\transport^{(i,j-1)}_{\link{x_1 x_2}}
	\circ\transport^{(i,j-1)}_{\link{x_2 x_3}} \neq -1$$ 
	Choose $M$ and $N$ to be even integers.
	The Chern number of the lattice bundle $\xi_{\Lambda(M,N)}$ is
	then given by

	\begin{equation}\label{e-t2plaqangle}
	\Ch(\xi_{\Lambda}) = -\frac{\I}{2\pi}
		\sum_{\lpoint{i}{j}}
		\log \left(
			\transport^{(i,j)}_{\link{x_2 x_1}}
			\circ\transport^{(i,j-1)}_{\link{x_4 x_1}}
			\circ\transport^{(i,j-1)}_{\link{x_1 x_2}}  
			\circ\transport^{(i,j-1)}_{\link{x_2 x_3}}			
			\right).
	\end{equation}

\end{Theorem}
\begin{proof}
	Previous calculation.
\end{proof}

	Note that such configurations for which the above Theorem 
	holds are often called \EM{continuous configurations} and the excluded
	ones are called \EM{exceptional configurations}.

	\begin{figure}
	\begin{center}	
		\psfig{figure=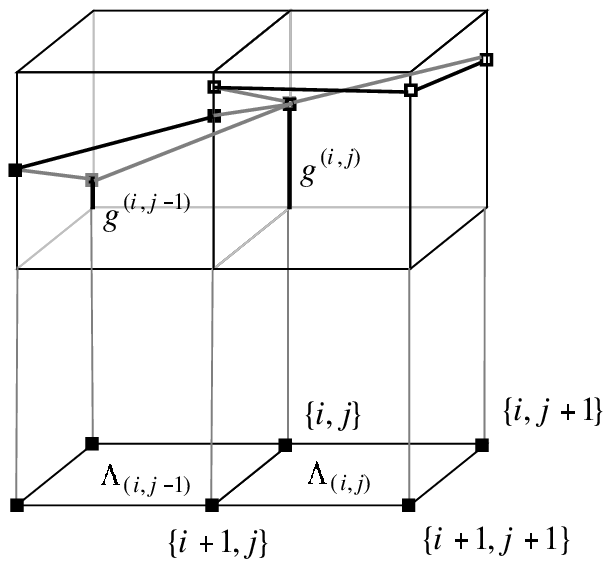}
	\end{center}
		\caption{Two neighboring local trivialisations}
		\protect\label{f-t2gij}
	\end{figure}

	If we denote the lattice parallel translations
	according to the standard notation 
	in lattice field theories,
	i.e.
	\begin{eqnarray*}
		U_{\lpoint{i}{j-1},\hat 1} & := &\transport^{(i,j-1)}_{\link{x_1 x_2}},\\  
		U_{\lpoint{i}{j-1},\hat 2} & := &\transport^{(i,j-1)}_{\link{x_1 x_4}},\\	
		U_{\lpoint{i+1}{j-1},\hat 2} & := &\transport^{(i,j-1)}_{\link{x_2 x_3}},\\	
		U_{\lpoint{i}{j},\hat 1} & := &\transport^{(i,j)}_{\link{x_1 x_2}},			
	\end{eqnarray*}	
	we obtain for (\ref{e-t2plaqangle})
	$$
		\Ch(\xi_{\Lambda}) = -\frac{\I}{2\pi}
		\sum_{\lpoint{i}{j}}
		\log \left({U_{\lpoint{i}{j},\hat 1}}\inv
			\circ {U_{\lpoint{i}{j-1},\hat 2}}\inv
			\circ U_{\lpoint{i}{j-1},\hat 1} 
			\circ U_{\lpoint{i+1}{j-1},\hat 2}\right),
	$$
	where the logarithm
	$$
		K_{(i,j-1)}:= \frac{\I}{2\pi}
		\log \left({U_{\lpoint{i}{j},\hat 1}}\inv
			\circ {U_{\lpoint{i}{j-1},\hat 2}}\inv
			\circ U_{\lpoint{i}{j-1},\hat 1} 
			\circ U_{\lpoint{i+1}{j-1},\hat 2}\right),
	$$
	is called the \EM{plaquette angle} of the plaquette
	$\Lambda_{(i,j-1)}$ and corresponds to the result obtained in \cite{Ph84}.

\section{Summary}

 Starting with the physically reasonable assumption of a connection
which is locally represented by pure gauges, we were basically able to
calculate or better to assign a Chern number to each \EM{configuration
on $\Lambda$}. This so obtained result is unfortunately not consistent
with the usual understanding of lattice gauge invariance. However even
more problematic is the fact that the general result for
$\Ch(\xi_{\Lambda})$ does not lead to $-\Ch(\xi_{\Lambda})$ for all
\EM{configurations on $\Lambda$} when 
inverting all parallel translations $\transport_{\link{x y}}$. These two
problems can be resolved with one additional assumption on the
connection which is expressed in an assumption on the parameterization
of the transition functions such that the integrals over the overlap
areas are less than $\pi$. This can always be assumed as far as
$\transport^{(i,j)}_{\link{x_2 x_1}}
\circ\transport^{(i,j-1)}_{\link{x_4 x_1}}
\circ\transport^{(i,j-1)}_{\link{x_1 x_2}}
\circ\transport^{(i,j-1)}_{\link{x_2 x_3}} \neq -1$ for all $\{i,j\}$.
As already observed in \cite{Ph84} without such a condition or at
least some restricting assumption there is no unique result.  Depending
on the parameterization of $\U1$ there is always one group element which,
to put it crudely, allows for \EM{two results} thus a tie breaker is
needed.

\section*{Acknowledgments}
\markboth{Acknowledgments}{}

We would like to thank Ch. Gattringer, H. Grosse, C.B. Lang  and L. Pittner
for many discussions.

%
%
\cleardoublepage
{\makeatletter\let\cleardoublepage\clearpage\let\chaptermark\@gobble

}


\begin{thebibliography}{10}

\bibitem{Fr92}
M.~P. Fry,
\newblock Phys. Rev. D 45 (1992) 682.

\bibitem{Fr93}
M.~P. Fry,
\newblock Phys. Rev. D 47 (1993) 2629.

\bibitem{GaSe94}
C.~R. Gattringer and E. Seiler,
\newblock Ann. Phys. 233 (1994) 97.

\bibitem{GaLa94b}
H. Gausterer and C.~B. Lang,
\newblock Phys. Lett. B 341 (1994) 46.

\bibitem{AzDiGa94}
V. Azcoiti, G.~D. Carlo, A. Galante, A.~F. Grillo, and V. Laliena,
\newblock Phys. Rev. D 50 (1994).

\bibitem{Fr95}
M.~P. Fry,
\newblock Phys. Rev. D 51 (1995) 810.

\bibitem{GaLa95}
H. Gausterer and C.~B. Lang,
\newblock Nucl. Phys. B 455 (1995) 785,

\bibitem{Ga95}
C. Gattringer,
\newblock {\em $QED_2$ and $U(1)$-Problem},
\newblock PhD thesis, Karl-Franzens-Universit{\"a}t Graz, Austria, 1995,
\newblock hep-th/9503137.

\bibitem{Ga95a}
C. Gattringer,
\newblock MPI-Ph/92-52, 1995.

\bibitem{AzDiGa96b}
V. Azcoiti, G.~D. Carlo, A. Galante, A.~F. Grillo, and V. Laliena,
\newblock Phys. Rev. D 53 (1996) 5069,

\bibitem{IrSe96}
A.~C. Irving and J.~C. Sexton,
\newblock Nucl. Phys. (Proc. Suppl.) 47 (1996) 679.

\bibitem{Sc62a}
J. Schwinger,
\newblock Phys. Rev. 125 (1962) 397.

\bibitem{Sc62b}
J. Schwinger,
\newblock Phys. Rev. 128 (1962) 2425.

\bibitem{Ad69}
S.~L. Adler,
\newblock Phys. Rev. 177 (1969) 2426.

\bibitem{BeJa69}
J.~S. Bell and R. Jackiw,
\newblock Nuovo Cim. 60 (1969) 47.

\bibitem{Wi79}
E. Witten,
\newblock Nucl. Phys. B156 (1979) 269.

\bibitem{Ve79}
G. Veneziano,
\newblock Nucl. Phys. B 159 (1979) 213.

\bibitem{BaAl96}
A.~P. Balachandran et~al.,
\newblock ESI 299  (1996).

\bibitem{GrMa92}
H. Grosse and J. Madore,
\newblock Phys. Lett. B 283 (1992) 218.

\bibitem{GrKl96}
H. Grosse, C. Klim{\v c}ik, and P. Pre{\v s}najder,
\newblock Commun. Math. Phys. 178 (1996) 507.

\bibitem{Lu82a}
M. L{\"u}scher,
\newblock Commun. Math. Phys. 85 (1982) 39.

\bibitem{Ph84}
A. Phillips,
\newblock Ann. Phys. (N.Y.) 161 (1985) 399.

\bibitem{AtSi68}
M. Atiyah and I. Singer,
\newblock Ann. Math. 87 (1968) 484.

\bibitem{SeSt82}
E. Seiler and I.~O. Stamatescu,
\newblock Phys. Rev. D25 (1982) 2177.

\bibitem{SmVi87}
J. Smit and J.~C. Vink,
\newblock Nucl. Phys. B 284 (1987) 234.

\bibitem{SmVi87a}
J. Smit and J.~C. Vink,
\newblock Nucl. Phys. B 286 (1987) 485.

\bibitem{Vi88}
J.~C. Vink,
\newblock Nucl. Phys. B Proc. Suppl. 4 (1988) 519.

\end{thebibliography}
\end{document}